\documentclass[fleqn,usenatbib]{mnras}

\usepackage{newtxtext,newtxmath}

\usepackage[T1]{fontenc}

\DeclareRobustCommand{\VAN}[3]{#2}
\let\VANthebibliography\thebibliography
\def\thebibliography{\DeclareRobustCommand{\VAN}[3]{##3}\VANthebibliography}

\usepackage{graphicx}
\usepackage{amsmath}
\usepackage{physics}
\usepackage{subcaption}
\usepackage{bm, upgreek}

\usepackage{soul}
\usepackage[dvipsnames]{xcolor}
\newcommand{\eplg}{EPL$\gamma$}

\newcommand{\condprob}[2]{\mathrm{Pr}\left(#1\vert#2\right)}
\newcommand{\prob}[1]{\mathrm{Pr}\left(#1\right)}
\newcommand{\data}{\mathbf{d}}
\newcommand{\massmodel}{\bm{\uptheta}_\mathrm{MM}}
\newcommand{\mamps}{\bm{\uptheta}_m}

\title[Multipoles account for flux-ratio anomalies]{General multipoles and their implications for dark matter inference}

\author[J. S. Cohen et al.]{
Jacob S. Cohen,$^{1}$\thanks{E-mail: jascohen@ucdavis.edu}
Christopher D. Fassnacht,$^{1}$
Conor M.~O'Riordan$^{2}$ and
Simona Vegetti$^{2}$
\\
$^{1}$Department of Physics and Astronomy, University of California Davis, 1 Shields Avenue, Davis, CA 95616, USA\\
$^{2}$Max Planck Institute for Astrophysics, Karl-Schwarzschild Str. 1, D-85748 Garching bei M\"unchen, Germany
}

\date{Accepted XXX. Received YYY; in original form ZZZ}

\pubyear{2024}

\begin{document}
\label{firstpage}
\pagerange{\pageref{firstpage}--\pageref{lastpage}}
\maketitle

\begin{abstract}
The flux ratios of strongly lensed quasars have previously been used to infer the properties of dark matter. In these analyses it is crucial to separate the effect of the main lensing galaxy and the low-mass dark matter halo population. In this work, we investigate flux-ratio perturbations resulting from general third- and fourth-order multipole perturbations to the main lensing galaxy's mass profile. We simulate four lens systems, each with a different lensing configuration, without multipoles. The simulated flux ratios are perturbed by 10-40 per cent by a population of low-mass haloes consistent with CDM and, in one case, also a satellite galaxy. This level of perturbation is comparable to the magnitude of flux-ratio anomalies in real data that has been previously analyzed. We then attempt to fit the simulated systems using multipoles instead of low-mass haloes. We find that multipoles with amplitudes of 0.01 or less can produce flux-ratio perturbations in excess of 40 per cent. In all cases, third- or fourth-order multipoles can individually reduce the magnitude of, if not eliminate, flux-ratio anomalies. When both multipole orders are jointly included, all simulated flux ratios can be fit to within the observational uncertainty. Our results indicate that low-mass haloes and multipoles are highly degenerate when modelling quadruply-imaged quasars based just on image positions and flux ratios. In the presence of this degeneracy, flux-ratio anomalies in lensed quasars alone cannot be used to place strong constraints on the properties of dark matter without additional information that can inform our priors.

\end{abstract}


\section{Introduction}
\label{sec:intro}

The cold dark matter (CDM) paradigm, which posits that dark matter consists of non-relativistic, collisionless particles \citep[e.g.][]{planck2016}, is successful at describing cosmic structure at scales larger than $\sim1$ Mpc \citep{springel2005, planck2020} and has been adopted as the standard in cosmology. CDM predicts a clumpy distribution of dark matter on sub-galactic scales and the existence of a large population of low-mass haloes \citep[e.g.,][]{vogelsberger2014, schaye2015}. On the other hand, warm dark matter models (WDM) predict a smaller amount of such objects with a less concentrated mass density profile \citep[e.g.][]{bode2001, viel2005, lovell2014, lovell2020}. The difference between CDM and still-viable WDM models is strongest at halo masses lower than $10^9 M_\odot$ \citep{hsueh2020}, where most of these objects are predicted to be faint or even completely dark. Strong gravitational lensing allows us to detect them via their gravitational effect on the strongly lensed images. 

In this paper, we focus on galaxy-scale strong lensing of unresolved sources, specifically quadruply-imaged quasars (quads). The image configuration is determined by the mass distribution of the lens and the position of the source. Low-mass haloes associated with the lens galaxy, called subhaloes, and haloes along the line of sight, called field haloes, can produce measurable changes in the relative fluxes of the lensed images due to the dependence of the image magnifications on the second derivative of the lensing potential. 

This method of investigating dark matter, known as flux-ratio analysis, cannot be used to precisely determine the masses and positions of individual haloes that cause the perturbations. Rather, the viability of a dark matter model is assessed based on the probability that its halo population could have produced the observed flux ratios, marginalised over the possible individual halo configurations. Dark matter models that produce larger numbers of low-mass haloes will lead to a higher incidence of lens systems that show so-called flux-ratio anomalies, in which a standard unperturbed smooth mass distribution cannot reproduce the observed flux ratios.  In contrast, dark matter models that suppress the formation of low-mass haloes will lead to fewer and less significant flux-ratio anomalies in samples of lensed quasar systems.  Single lens systems provide only a weak inference on dark matter models, since even in the presence of a large number of associated low-mass perturbers, these may be spatially distributed in a way such that no flux-ratio anomaly is produced.  Thus, large samples of lens systems are needed.

Flux-ratio analysis was first proposed by \cite{mao1998} and \cite{metcalf2001}, and it was originally limited to subhaloes. Soon after, \cite{dalal2002} applied it to a sample of seven quads and reported results that were consistent with CDM simulations at 90 per cent confidence. 

Follow-up studies argued for the inclusion of additional components that could also influence flux ratios, such as field haloes and stellar disks \citep{moller2003, inoue2012, metcalf2005, despali2018}. The contribution from field haloes is especially important given that they are often more numerous than substructures \citep{despali2018}. Furthermore, field haloes should provide a cleaner test of dark matter models because their properties are not influenced by the tidal effects that can be so important for subhaloes.  Baryonic components such as stellar disks have been discovered in real lens systems and, when included in the lens model, have successfully been able to reproduce the flux-ratio anomalies in those systems without having to resort to low-mass dark matter haloes \citep{hsueh2016,hsueh2017}.  Similarly, more general explorations have shown that baryonic structures in lensing galaxies can mimic perturbations by low-mass haloes if not properly accounted for in the lens model \citep{gilman2017,gilman2018,hsueh2018}. Sensitive high-resolution imaging can be used to estimate the contribution of baryonic structure in lensing galaxies, but other complexities may remain as confounding factors.

In this paper, we will investigate an important form of additional complexity for lens mass models, namely, the angular structure in the lensing galaxy, parameterised here as multipole perturbations.  The most common mass profile used to model lens galaxies is the elliptical power law (EPL) with external shear \citep[e.g.,][]{tessore2015}.  We will hereafter refer to this type of base model as the \eplg\ model.  The multipoles that we consider add Fourier-type perturbations to the angular part of the density profile, leaving the radial part unchanged.  The use of multipoles is motivated by optical and infrared observations of elliptical galaxies, which show that the isophotes of many of them deviate from perfect ellipticity \citep[e.g.,][]{bender1988, bender1989, cappellari2016}.  These deviations can be modeled by simply-parameterized multipole components.  While many treatments of elliptical galaxy isophotes focus only on fourth-order multipoles, \citet{hao2006} present an extensive investigation of the surface brightness distribution in elliptical galaxies in which they fit both third- and fourth-order multipoles with a variety of orientation angles. 

It is thus natural to consider multipole components in the mass distributions of galaxies as well.  Demonstrating the impact of angular complexity in the lens galaxy on flux ratios, \cite{evans2003} and \cite{cogndon2005} showed that the joint inclusion of third- and fourth-order multipoles with unrestricted orientation angles can reproduce many anomalies that had been observed at the time. We focus specifically on third- and fourth-order multipoles because they are expected to cause flux-ratio effects degenerate with those of perturbing haloes. Lower-order multipoles have effects that are analogous to changes in the macro-model parameters, and higher-order multipoles may introduce deviations from ellipticity that either produce greater than four images or are unphysical \citep{evans2003, cogndon2005}. Despite these findings, flux-ratio analyses since then have neglected to implement them completely. \cite{hsueh2020} study some of the same lenses as \cite{evans2003} and \cite{cogndon2005}, including angular structure in one lens in the form of an exponential disk, but they modelled all other lenses with only an EPL and shear. Recent flux-ratio analyses by \cite{gilman2021, gilman2022, gilman2023} have included multipoles but in a specific and restrictive way. In those analyses, the orientation angle of the fourth order multipole is fixed to align with the EPL. Third order multipoles are not included. \cite{gilman2024} inferred constraints on WDM from simulated lens systems including both third- and fourth-order multipoles, but the fourth order was fixed to align with the EPL. The effect of multipoles has also been considered in the context of the extended emission of lensed galaxies, where \cite{oriordan2023} found that the inclusion of third- and fourth-order multipoles with unrestricted orientation angles could produce false substructure detections. 
 
In this paper, we extend  earlier work on lensed quasar flux-ratios \citep{evans2003,cogndon2005} in several important ways.  First, while those papers modeled real lenses with multipole components, our investigation uses simulated lenses so that we can directly compare the perturbative effects of low-mass haloes with those of multipoles. In addition, we consider the effects of third- and fourth-order multipoles separately as well as jointly, and have more generality in our base models by allowing the power-law index to be different from the isothermal value. Our particular focus is an investigation of the potential for general third- and fourth-order multipoles to perturb the flux ratios of quadruply-imaged quasars in a way that is degenerate with perturbations from low-mass dark matter haloes. In Section \ref{sec:real}, we describe our procedure for obtaining \eplg\ base models for a sample of real lens systems. In Section \ref{sec:mocks}, we detail the creation of four simulated lens systems from the combination of \eplg\ base models and CDM low-mass halo populations plus, for one of the systems, a satellite galaxy.  Section \ref{sec:sim} describes how we model the simulated lenses in our sample using just an \eplg\ model plus multipole components, and Section \ref{sec:results} presents the results. We discuss the implications of our results and future work in Section \ref{sec:conc}.

\section{Modelling real data: \eplg}
\label{sec:real}

To quantify the effect of multipoles in realistic scenarios, we create simulated strong gravitational lens systems using image configurations from real lens systems taken from recent flux-ratio analysis studies \citep{hsueh2020, gilman2020}. To ensure applicability of our results across different image configurations, we select lens systems that fall into one of each of the general categories (see Figure \ref{fig:fig1} for visualizations): cross (WGD J0405-3308), fold (WFI 2026-4536) and cusp (B1422+231). We also select the cross configuration system PS J1606-2333, which has a luminous satellite associated with the main lensing galaxy, as the basis for a fourth simulated strong lens system. The satellite will allow us to investigate the degeneracy between multipoles and haloes beyond the low-mass range we otherwise consider (see Section \ref{sec:perturbers}).

For the mass model, we use an \eplg. The corresponding dimensionless surface mass density (convergence) is given by
\begin{equation}
    \kappa(R) = \frac{2-t}{2} \left(\frac{b}{R}\right)^t \qq{,}
    \label{pl_kappa}
\end{equation}
where $R$ is an ellitpical radius such that $R^2=(qx)^2+y^2$. The model parameters are the power-law slope, $t$, axis ratio, $q$, and scale length, $b=R_E\sqrt{q}$, where $R_E$ is the Einstein radius. We define external shear with amplitude, $\gamma_{\rm ext}$, and orientation angle, $\phi_{\rm ext}$. The satellite in J1606 is modelled as a singular isothermal sphere (SIS), an EPL profile with $t=1$ and $q=1$. 

Fitting only the observed image positions, we perform Markov chain Monte Carlo (MCMC) sampling to approximate posterior distributions of the mass model parameters for each lens system. The location of the luminous satellite in J1606 is fixed to the observed position, but we allow all other parameters, including the source position, to vary freely. For convenience, we will hereafter refer to the distributions generated in this step, including the system with the satellite, as \eplg\ distributions. These sets of parameters describe the base models of our simulated data that will be perturbed either by low-mass haloes (\eplg$+$CDM; Section~\ref{sec:mocks}) or by multipoles (\eplg$+$MP; Section~\ref{sec:sim}). 

\begin{table*}
	\centering
	\caption{Macro-model parameters used to create mock observations. The columns are $z_l$, the lens redshift, $z_s$, the source redshift, $\theta_{\rm E}$, the Einstein radius, dRA, the lens right ascension with respect to the observation center, dDec, the lens declination with respect to the observation center, $t$, the power law slope, $\epsilon$, the ellipticity, $\phi$, the position angle, $\gamma_{\rm ext}$, the external shear strength, $\phi_{\rm ext}$, the shear angle, dRA$_{\rm src}$, the source right ascension with respect to the observation center, and dDec$_{\rm src}$, the source declination with respect to the observation center.}
	\label{tab:macro}
	\begin{tabular}{lcccccccccccc} 
		\hline
		Lens & $z_l$ & $z_s$ & $\theta_{\rm E}$ & dRA & dDec & $t$ & $\epsilon$ & $\phi$ & $\gamma_{\rm ext}$ & $\phi_{\rm ext}$ & dRA$_{\rm src}$ & dDec$_{\rm src}$\\
		\hline
		Cross & 0.29 & 1.713 & 0.702 & 0.380 & 0.552 & 1.013 & 0.076 & -62.66 & 0.031 & -2.89 & 0.385 & 0.560\\
		Fold & 1.04 & 2.23 & 0.655 & -0.003 & 0.000 & 1.00 & 0.100 & -90.00 & 0.100 & 90.00 & 0.030 & 0.079\\
  		Cusp & 0.34 & 3.62 & 0.765 & 0.699 & -0.629 & 1.00 & 0.176 & 60.90 & 0.193 & 52.96 & 0.388 & -0.414 \\
		  Satellite (main) & 0.31 & 1.70 & 0.653 & 0.794 & 0.248 & 1.02 & 0.140 & -54.51 & 0.216 & 26.81 & 0.798 & 0.177 \\
		Satellite (satellite) &  &  & 0.101 & 0.337 & -1.032 & 1.00 & 0.0 & & & & & \\[.1cm]
		\hline
	\end{tabular}
\end{table*}

\section{Simulated data: \eplg$+$CDM}
\label{sec:mocks}

We create our four simulated lens systems by adding populations of CDM subhaloes and field haloes to the base models drawn from the \eplg\ distributions described in Section \ref{sec:real}. The results are mock quads with image positions that match those of a real lens system and flux ratios that are perturbed only by low-mass haloes. These lens systems do not contain any multipole components. We do not include both low-mass haloes and multipoles in any of our simulated data or models, as we intend to investigate whether the same observations can be produced with the presence of either individually. This represents a conservative approach to study their degeneracy.

\subsection{Background source}
\label{sec:src}

Typically,  flux-ratio investigations focus on emission from regions of the background objects that are large enough to avoid being affected by microlensing by stars in the primary lensing galaxy. These include mid-infrared emission from dust surrounding quasar accretion disks, which are typically smaller than 10 pc \citep{burtscher2013}; emission from the narrow-line regions surrounding a quasar, which can extend up to 60 pc \citep{muller2011, nierenberg2017}; or radio emitting regions, for which  individual observations give estimates of sizes smaller than 10 pc \citep{lee2017, kim2022}. 
Generally, as the size of the background source increases, it becomes less susceptible to flux perturbations from low-mass haloes \citep{dobler2006}. In this paper, we want to quantify the degeneracy between low-mass haloes and multipoles in the scenario in which the effect of the former is maximal, hence the background sources in our mock observations and models are point-like. Their location in each realization is drawn from the MCMC chains associated with the modelling of the real data. In Section \ref{sec:conc}, we further investigate the effect of the source size and its implication for the degeneracy under study.

\subsection{Low-mass halo population}
\label{sec:perturbers}

To generate the CDM halo populations that we add to the \eplg\ models, we largely follow the process described in \cite{hsueh2020} with updated treatments of the mass-concentration relations for subhaloes and field haloes. All low-mass haloes are modelled as NFW profiles  (\citealp{NFW1997}; however, see \citealp{heinze2024}).

For the field halo mass-concentration relation, we use that reported in Table 1 of \cite{duffy2008} from N-body simulations. We use values derived using the virial radius definition of relaxed haloes between redshifts 0 and 2. Unlike \cite{hsueh2020}, we apply the associated scatter on the parameters. We follow the implementation of \cite{despali2016}, which is based on the approach introduced by \cite{sheth1999}, for the field halo mass function. We use their best-fitting parameters optimized over all considered redshifts and cosmologies.

We determine subhalo concentrations from a redshift-dependent mass-concentration relation extracted from the ShinUchuu N-body simulation (\citealp{ando2020, moline2023}, also see \citealp{conor2023} for more details). This relation is derived in terms of $R_{\rm max}$ and $m_{\rm max}$, the radius of maximum tangential velocity and mass enclosed within it, as these more accurately describe the characteristics of haloes in simulations than the usual virial quantities. Our choice of mass-concentration relation results in more concentrated subhaloes than does the typical one from \cite{duffy2008}. We use a subhalo mass function that comes from fitting to the data in \cite{lovell2020} and has been reparameterised in terms of $m_{\rm max}$. After drawing the subhalo mass, $m_{\rm max}$, from the mass function, we draw the corresponding $R_{\rm max}$ value from a log-normal distribution with mean
\begin{equation}
    R_{\rm max, mean}^{\rm CDM} = \left( \frac{A^{2/B} G m_{\rm max} M_\odot}{100\  {\rm (km/s)^{-2} \ kpc}\  M_\odot} \right)^{B/(B+2)} \rm kpc \qq{,}
	\label{eq:rmaxmeancdm}
\end{equation}
and standard deviation
\begin{equation}
   \sigma =  \exp \left[(p + mz) + (n - kz) \ln \left(\frac{m_{\rm max}}{10^{10} M_\odot}\right)\right] \qq{.}
	\label{eq:rmaxsigma}
\end{equation}
Here, $A=az+d$ and $B=cz+b$. Values for $a$, $b$, $c$, $d$, $k$, $m$, $n$, and $p$ are listed in Table \ref{tab:rmaxparams}, and $z$ is the redshift. The NFW profile for a subhalo then has normalization
\begin{equation}
    \rho_0[M_\odot {\rm kpc^{-3}}] = \frac{ m_{\rm max}(1+C)^2}{4 \pi C^2 R_{\rm s}^3} \qq{,}
	\label{eq:rho_s}
\end{equation}
where $R_{\rm s}$ is the scale radius and $C= R_{\rm max}/R_{\rm s}=2.16$ \citep{bullock2001}. We do not include tidal truncation or a dependence on the distance from the main lens centre of the mass-concentration relation because \cite{despali2018} have shown the effects to be small compared to the scatter on the mass-concentration relation.

We generate populations consistent with predicted CDM subhalo and field halo mass functions down to a halo mass of $~10^5 M_\odot$, and we assume the total mass in substructure in the region of the lensed images to be $\sim 2$ per cent of the total mass of the main lens in that region, which is roughly consistent with observational constraints \citep{dalal2002,hsueh2020,gilman2020}. This substructure fraction is higher than simulation predictions by \cite{xu2015}, but since we are testing the ability of multipoles to mimic low-mass haloes that strongly perturb the flux ratios, a bias towards models that have more perturbing haloes is a conservative choice.

\subsection{Selecting realizations for simulated lens systems}
\label{sec:selecting}

Because we will proceed to stress test multipoles as they try to reproduce the flux ratios resulting from these \eplg$+$CDM models, we generate 2000 realizations for each of the four main types of simulated lens in our sample (cross, fold, cusp, or satellite) and then select for each type the realization that produces the most extreme flux-ratio anomalies. These four \eplg$+$CDM models should thus present the flux ratios that are most difficult for the \eplg$+$MP models to reproduce. If models with multipoles but without low-mass haloes can fit perturbations produced by the most extreme halo populations, they should be able to do so in nearly all cases. We stress that though they are strong, the perturbations in our simulated lens systems are of comparable magnitude to flux-ratio anomalies in real observations \citep[e.g.][]{nierenberg2020}. Each simulated lens system contains total flux-ratio perturbations in excess of 10 per cent, and some images are perturbed beyond 20 per cent. Table \ref{tab:macro} lists the macro-model parameters used in each of the four simulated systems, and Table \ref{tab:mock_obs} presents the image positions and flux ratios. We add to each flux in the models uncertainties that are based on the observations of the real lenses on which they were based.

\begin{table}
	\centering
	\caption{Unitless constants used in equations \ref{eq:rmaxmeancdm} and \ref{eq:rmaxsigma}.}
	\label{tab:rmaxparams}
	\begin{tabular}{ccc} 
		\hline
		$R_{\rm max, mean}^{\rm CDM}$& \\
		\hline
		a & 0.24986592\\
		b & 1.55822031\\
		c & -0.01885084\\
		d & 0.38482671\\
		\hline
		$\sigma$& \\
		\hline
		p & 1.28099\\
		m & 0.21388\\
		n & 0.46263\\
		k & 0.01501\\
		\hline
	\end{tabular}
\end{table}

\section{Modelling of simulated data: \eplg$+$MP}
\label{sec:sim}

In accordance with previous works examining the lensing effects of complex angular structure, we describe the convergence of multipoles in polar form
\begin{equation}
    \label{multipole}
    \kappa(R, \phi) = R^{-t} [a_m \cos{(m\phi)} + b_m \sin{(m\phi)}] \qq{.}
\end{equation}
Here, $a_m$ and $b_m$ are the standard multipole sine and cosine amplitudes, and $m$ is the multipole order. For ease of interpretation, we also describe multipoles in terms of their overall amplitudes, 
\begin{equation}
    \label{eta_m}
    \eta_m = \sqrt{a_m^2 + b_m^2} \qq{,}
\end{equation}
and orientation angles,
\begin{equation}
    \label{phi_m}
    \phi_m = \frac{1}{m}\arctan\left(\frac{b_m}{a_m}\right) \in \frac{1}{m}[0,2\pi) \qq{.}
\end{equation} 
We restrict out focus in this paper to third- and fourth-order multipoles ($m = 3,4$). To assess the possible degeneracy between the lensing effects of multipoles and haloes above our standard mass range, we do not include the SIS satellite galaxy in any of our models of that lens system. 

\begin{table}
        \centering
	\caption{Image positions and flux ratios for each of the simulated lens systems along with the fractional degree of flux perturbation by low-mass haloes compared to the best-fitting EPL and shear-only fiducial model. All position units are in arcsec. These are generated using an EPL+shear macro-model in addition to a population of perturbing CDM subhaloes and field haloes. No multipoles are present in any of the simulated systems. Positions are given with respect to the observation centre, and all dRA and dDec uncertainties are $0.005$. Flux ratio uncertainties are listed in parentheses next to each flux ratio.}
	\label{tab:mock_obs}
	\begin{tabular}{llccc} 
		\hline
		Lens & Image & \multicolumn{2}{c}{Position} & Flux Ratio \\
            & & dRA & dDec & \\
		\hline
            Cross & A & 1.0656 & 0.3204 & 1.000 (0.030) \\
                  & B & 0.0026 & -0.0017 & 0.508 (0.001)\\
                  & C & 0.7222 & 1.1589 & 0.920 (0.030)\\
                  & D & -0.1562 & 1.0206 & 0.658 (0.030)\\
		\hline
            Fold & A1 & -0.4985 & -0.2207 & 0.288 (0.012)\\
                  & A2 & 0.2364 & -0.6048 & 1.000 (0.040)\\
                  & B & 0.4897 & -0.3895 & 0.893 (0.036)\\
                  & C & 0.0725 & 0.8233 & 0.299 (0.015)\\
		\hline
            Cusp & A & 0.3908 & 0.3213 & 1.000 (0.010)\\
                  & B & 0.0003 & 0.0003 & 1.149 (0.010)\\
                  & C & -0.3330 & -0.7463 & 0.537 (0.010)\\
                  & D & 0.9511 & -0.8018 & 0.045 (0.010)\\
		\hline
            Satellite & A & 1.6217 & 0.5890 & 0.867 (0.030)\\
                  & B & -0.0005 & 0.0003 & 1.000 (0.010)\\
                  & C & 0.8328 & -0.3170 & 0.670 (0.030)\\
                  & D & 0.4948 & 0.7377 & 0.694 (0.030)\\
		\hline
	\end{tabular}
\end{table}

\section{Results}
\label{sec:results}

In this section, we attempt to reproduce the simulated lens systems with multipoles in two distinct ways. First, we step through the parameter space of strengths and orientation angles for third- and fourth-order multipoles separately to determine the range of flux ratio perturbations that each order, strength and angle can produce (Section \ref{sec:indep}). In this first part, the multipoles are not fit to the data. Rather, we simply show that configurations of multipoles exist that can produce flux ratio anomalies consistent with those produced by low-mass haloes. Second, we then use MCMC to fit the multipole and mass model parameters to the flux ratios of the simulated systems. (Section \ref{sec:mcmc}).

\subsection{Independent investigation of third- and fourth-order multipoles}
\label{sec:indep}

At this stage, we are not trying to fit to the flux ratios in our simulated sample, but rather to explore the dependence of the flux-ratio perturbations on the multipole amplitudes ($\eta_m$, where $m$ is either 3 or 4)  and position angles ($\phi_m$).  We do this by generating a grid of ($\eta_m$, $\phi_m$) pairs and, for each grid point, adding multipole components with these parameters to the 200 base \eplg\ models. We do this exercise for $m=3$ and $m=4$ separately. While, judging from the isophotes, multipole amplitudes are not expected to be much larger than $\eta_m \approx 10^{-2}$ in real galaxies \citep[see][]{hao2006}, we explore 11 multipole strengths ranging from $10^{-3}-10^{-1}$ with equal logarithmic spacing. For each multipole strength, we examine an evenly spaced set of 10 orientation angles encompassing the full range over which they are unique, i.e., from $-$60 to $+$60 degrees for the third-order multipoles and from $-$45 to $+$45 degrees for the fourth-order multipoles.  To correct for any astrometric perturbations introduced by the addition of the multipole component, we optimize the macro-model parameters to fit to the simulated image positions after adding the multipole.

\begin{figure*}
\centering
    \includegraphics[width=\textwidth]{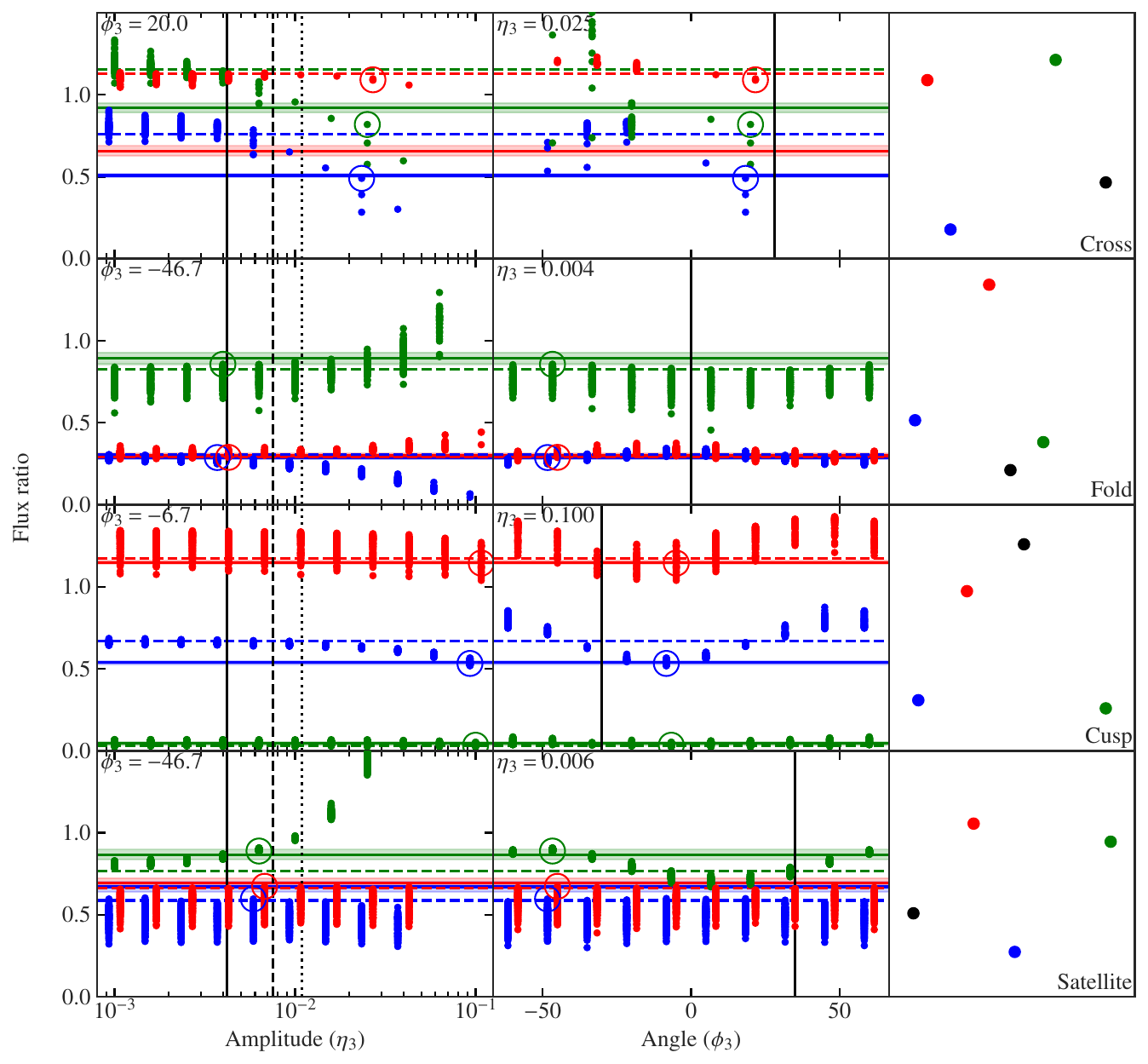}
    \caption{The range of flux-ratios as a function of third-order multipole parameters $\eta_3$ (left) and $\phi_3$ (middle). Only flux ratios from models with position error less than $1\sigma$ for each image are included. The horizontal lines and shaded regions represent the flux-ratios and uncertainties in the simulated data, which is generated by an \eplg$+$CDM model. The filled dots show the flux-ratios of each image for each multipole parameter value. Open circles represent the flux-ratios from the \eplg$+$MP model that most closely matches the simulated flux ratios. The solid, dashed and dotted vertical black lines in the left column indicate the mean, 1$\sigma$ and 2$\sigma$ values, respectively, of the third-order multipole amplitude distributions from elliptical galaxy isophotes \citep{hao2006}. The solid black line in the middle column indicates the orientation angle of the EPL major axis in each simulated lens system. The rightmost column shows the image configuration for each lens system, and the color of each image corresponds to data for its respective flux ratio in the other columns. The black point marks the image used in the denominator of the flux ratios.}
    \label{fig:fig1}
\end{figure*}

\begin{figure*}
\centering
    \includegraphics[width=\textwidth]{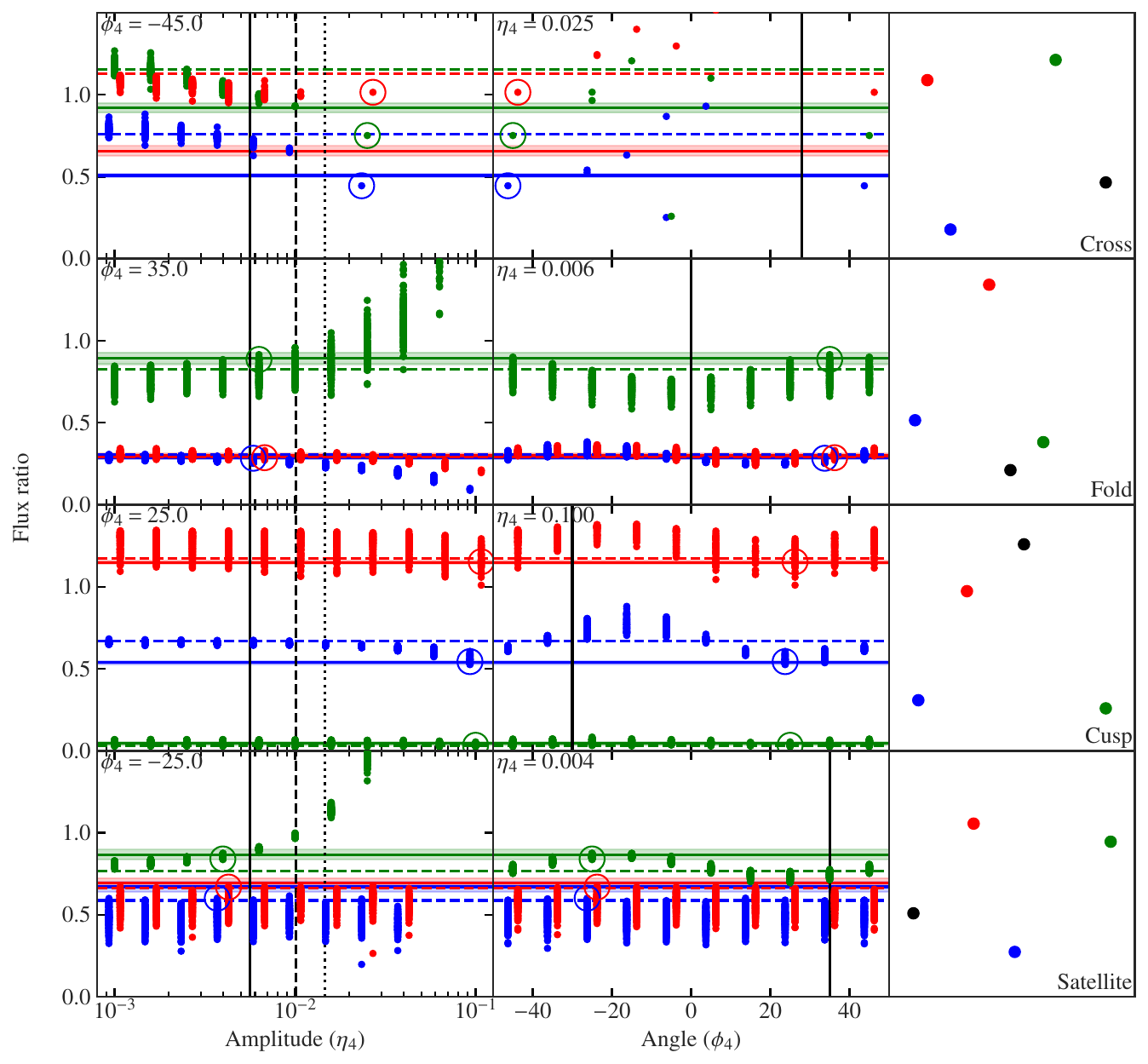}
    \caption{Same description as Figure \ref{fig:fig1}, except with fourth-order multipoles.}
    \label{fig:fig2}
\end{figure*}

We show the results of this exploration in Figures \ref{fig:fig1} (third-order) and \ref{fig:fig2} (fourth-order).  In each of the panels in the first two columns, the horizontal lines show the flux ratios of our simulated lenses while the points show the flux ratios produced by the \eplg$+$MP models. In the left-hand columns we show how the perturbations change with multipole amplitude, showing only the points for values of $\phi_m$ corresponding to the orientations of the highest-likelihood realizations. We calculate the likelihood of a realization from 
\begin{equation}
    \label{eq:chi_squared}
    \chi^2 = \sum_i \frac{\norm{\mathbf{x}^{\rm m}_i - \mathbf{x}^{\rm d}_i}^2}{\delta x_i^2} + \sum_i \frac{\left(f_i^{\rm m} -  f_i^{\rm d}\right)^2}{\delta f_i^2} \qq{,}
\end{equation}
where $\mathbf{x}_i$ and $f_i$ are the image positions and flux ratios of a model realization (denoted $^m$) and simulated data (denoted $^d$). $\delta x_i$ and $\delta f_i$ are the simulated uncertainties. 
As expected, flux-ratio perturbations get larger with increasing multipole amplitude. However, the size of this effect is dependent on the configuration and the particular image in question. 

The centre columns of Figures \ref{fig:fig1} and \ref{fig:fig2} show perturbations due to multipoles over the full range of orientation angles with $\eta_m$ values fixed to those of the highest-likelihood realizations. All four lens systems show clear periodic behavior as the orientation angle changes. In all cases, some realizations bring the flux ratios closer to the simulated values than the macro-model alone. The simulated flux ratios in the fold and cusp lens systems can be reproduced within observational uncertainty by either third- or fourth-order multipoles, though the highest-likelihood realizations for the cusp system both have potentially unrealistic\footnote{What we mean here by unrealistic is large compared to what is on average observed in the isophotes of elliptical galaxies. However, as we discuss in Section \ref{sec:conc}, the amplitude of multipole components in the light and mass distribution do not necessarily have to agree.} amplitudes of $\eta_m = 0.1$. Though no third- or fourth-order multipole perturbations can reproduce the simulated flux ratios in the cross and satellite systems, the magnitude of the discrepancy between model-predicted and simulated flux ratios can be significantly reduced by either order with reasonable amplitudes. 

The flux-ratio perturbations induced by the SIS in conjunction with low-mass haloes in our simulated satellite system are comparable to those induced by low-mass haloes alone in other mocks. While satellite galaxies may be directly observable from their light, there may be a degeneracy between their inferred properties and multipole amplitudes. We leave the investigation of this potential degeneracy to a future work.

\subsection{Joint investigation of third- and fourth-order multipoles}
\label{sec:mcmc}

We now fit the mass model and multipole parameters simultaneously to both the flux ratios and image positions of our simulated lens systems. These data are the elements of a vector $\data$. Similarly, the macro model parameters are the elements of a vector $\massmodel=\{\theta_E,...\}$, and the multipole amplitudes form a vector $\mamps=\{a_3,b_3,a_4,b_4\}$. From Bayes' theorem, the posterior distribution of the parameters given the data is
\begin{equation}
    \condprob{\mamps,\massmodel}{\data}=\frac{\condprob{\data}{\mamps,\massmodel}\prob{\mamps,\massmodel}}{\prob{\data}}.
\end{equation}
The first term in the numerator $\condprob{\data}{\mamps,\massmodel}$ depends only on $\chi^2$ defined previously (see Equation \ref{eq:chi_squared}). The prior probability $\prob{\mamps,\massmodel}=\prob{\mamps}\prob{\massmodel}$ is the probability of a given parameter value before the data is observed, based on other information. The normalisation of the posterior, or the evidence, $\prob{\data}$ can be ignored in this case as we only consider one model. The posterior we use in practice is then
\begin{equation}
    \condprob{\mamps,\massmodel}{\data}\propto\prob{\mamps,\massmodel}\exp\left(-\frac{1}{2}\chi^2\right).
\end{equation}
The calculation of this many-dimensional posterior is intractable, so we use MCMC to obtain samples of $\mamps$ and $\massmodel$. The density of these samples represents the posterior probability distribution. To conduct the MCMC sampling, we use the {\sc emcee}\footnote{\cite{emcee}} ensemble sampler with 120 walkers and 10,000 burn-in steps that are discarded, followed by 20,000 recorded steps. We use broad uniform priors on the mass model parameters $\massmodel$ and use a normally distributed prior with mean zero and standard deviation of one per cent on the multipole amplitudes.
\begin{figure*}
\centering
\begin{subfigure}{\columnwidth}
    \includegraphics[width=\columnwidth]{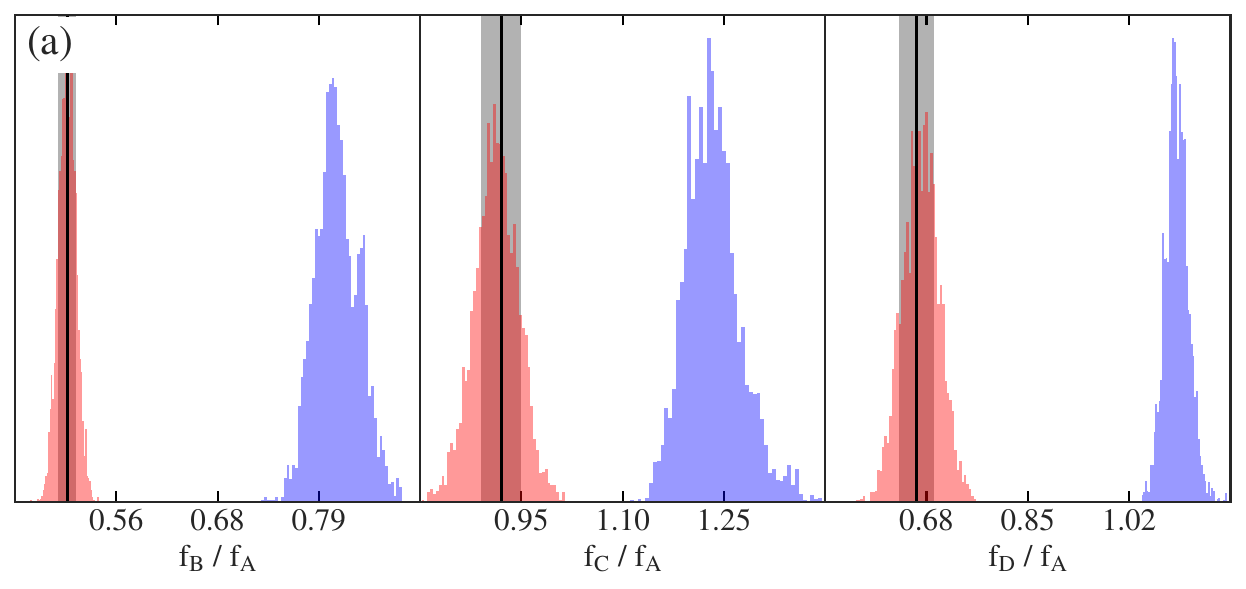}
    \label{fig:J1_flux_hists}
\end{subfigure}
\hfill
\begin{subfigure}{\columnwidth}
    \includegraphics[width=\columnwidth]{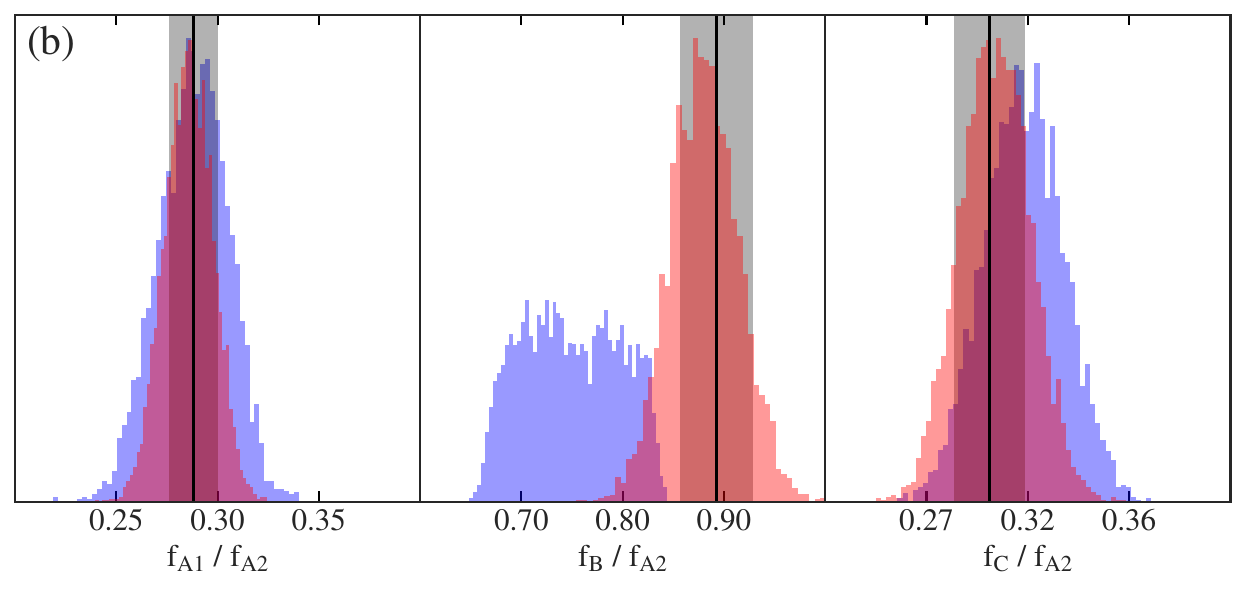}
    \label{fig:M1_flux_hists}
\end{subfigure}
\hfill
\begin{subfigure}{\columnwidth}
    \includegraphics[width=\columnwidth]{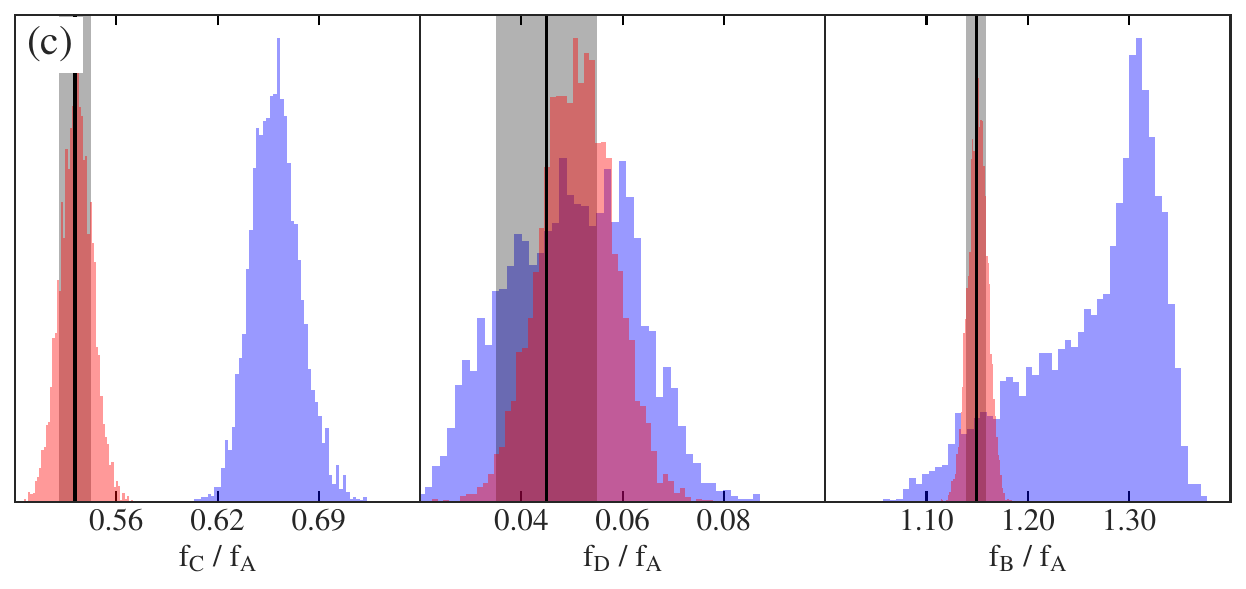}
    \label{fig:B1422_N_flux_hists}
\end{subfigure}
\hfill
\begin{subfigure}{\columnwidth}
    \includegraphics[width=\columnwidth]{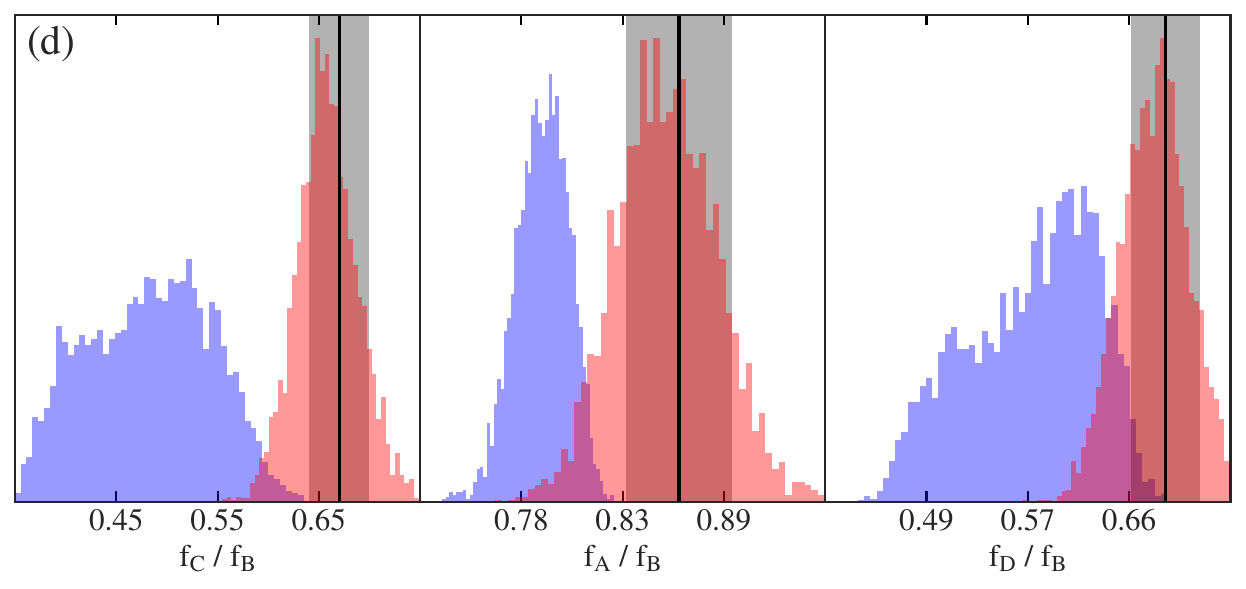}
    \label{fig:S1_flux_hists}
\end{subfigure}
\caption{Flux ratio histograms created using random samples of 10,000 models from MCMC chains for all four simulated lens systems: cross (a), fold (b), cusp (c) and satellite (d). For comparison, we show distributions resulting from using \eplg\ models (blue) and \eplg$+$MP models with third- and fourth-order multipoles included simultaneously (red). Vertical black lines show flux ratios from the simulated data, and the surrounding grey regions show the associated uncertainties.}
\label{fig:fig3}
\end{figure*}

\begin{figure*}
\centering
\begin{subfigure}{\columnwidth}
    \includegraphics[width=\columnwidth]{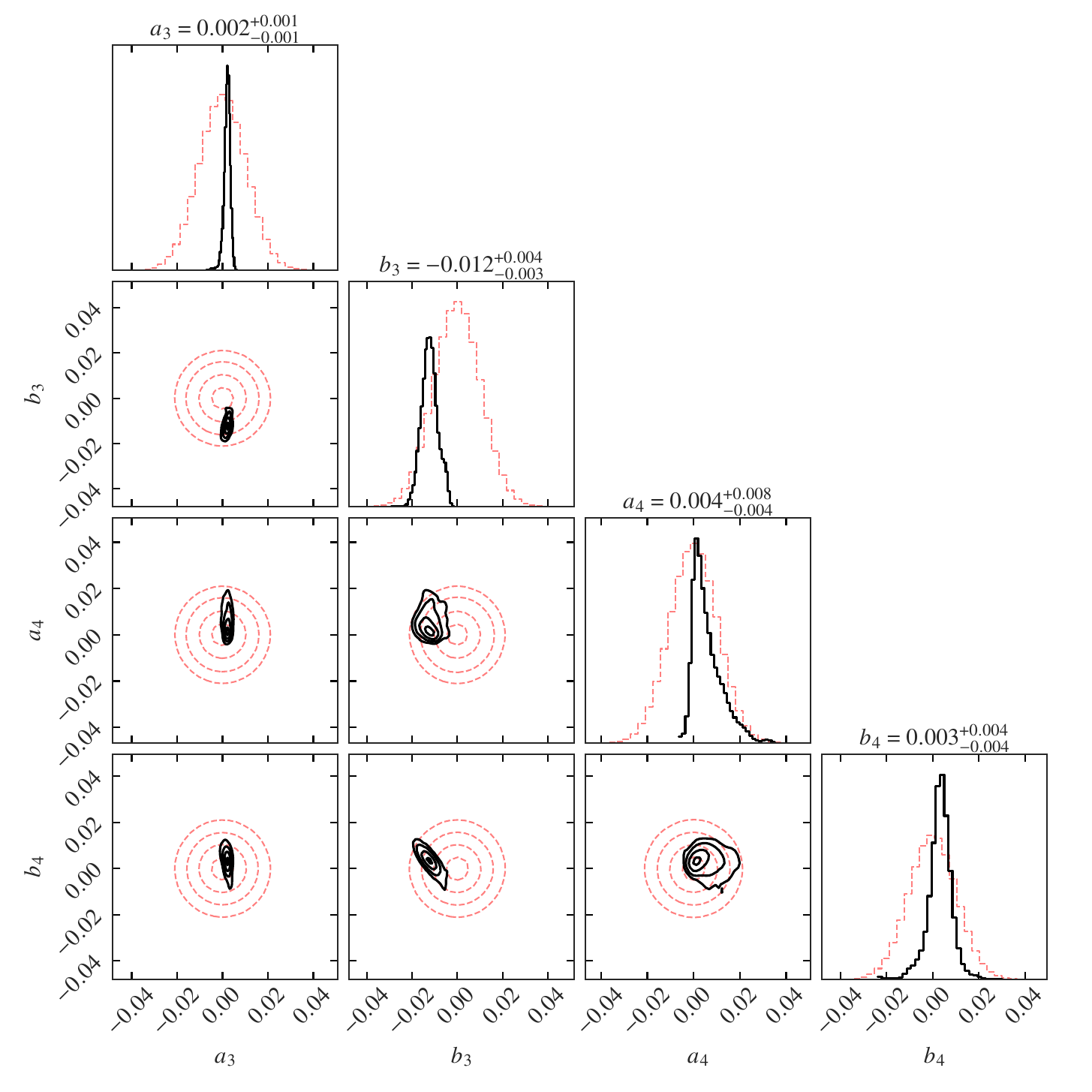}
    \label{fig:J1_eta_phi}
\end{subfigure}
\hfill
\begin{subfigure}{\columnwidth}
    \includegraphics[width=\columnwidth]{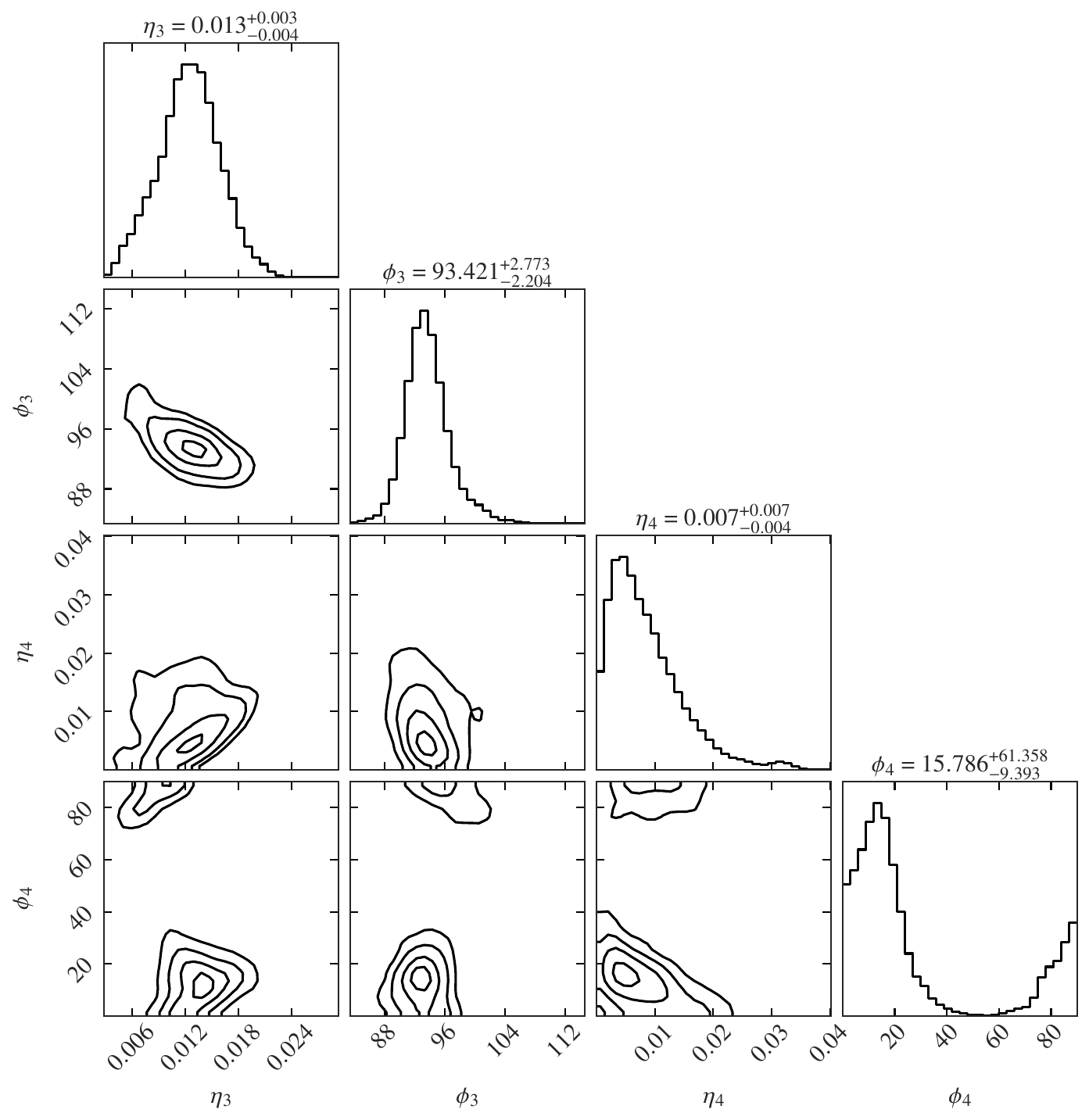}
    \label{fig:J1_amps}
\end{subfigure}
\caption{Corner plots from simultaneous MCMC sampling of EPL, shear and both multipole order parameters on the cross lens system. MCMC was conducted in the sine/cosine amplitude basis (left), and we also present the same data transformed into the overall amplitudes and angles (right) for ease of interpretation. Only multipole parameters are displayed, but all EPL and shear parameters, along with the source position, were allowed to vary freely. The red-dashed contours and histograms show the Gaussian prior distributions used for the $a_m$ and $b_m$ parameters. }
\label{fig:fig4}
\end{figure*}

\begin{figure*}
\centering
\begin{subfigure}{\columnwidth}
    \includegraphics[width=\columnwidth]{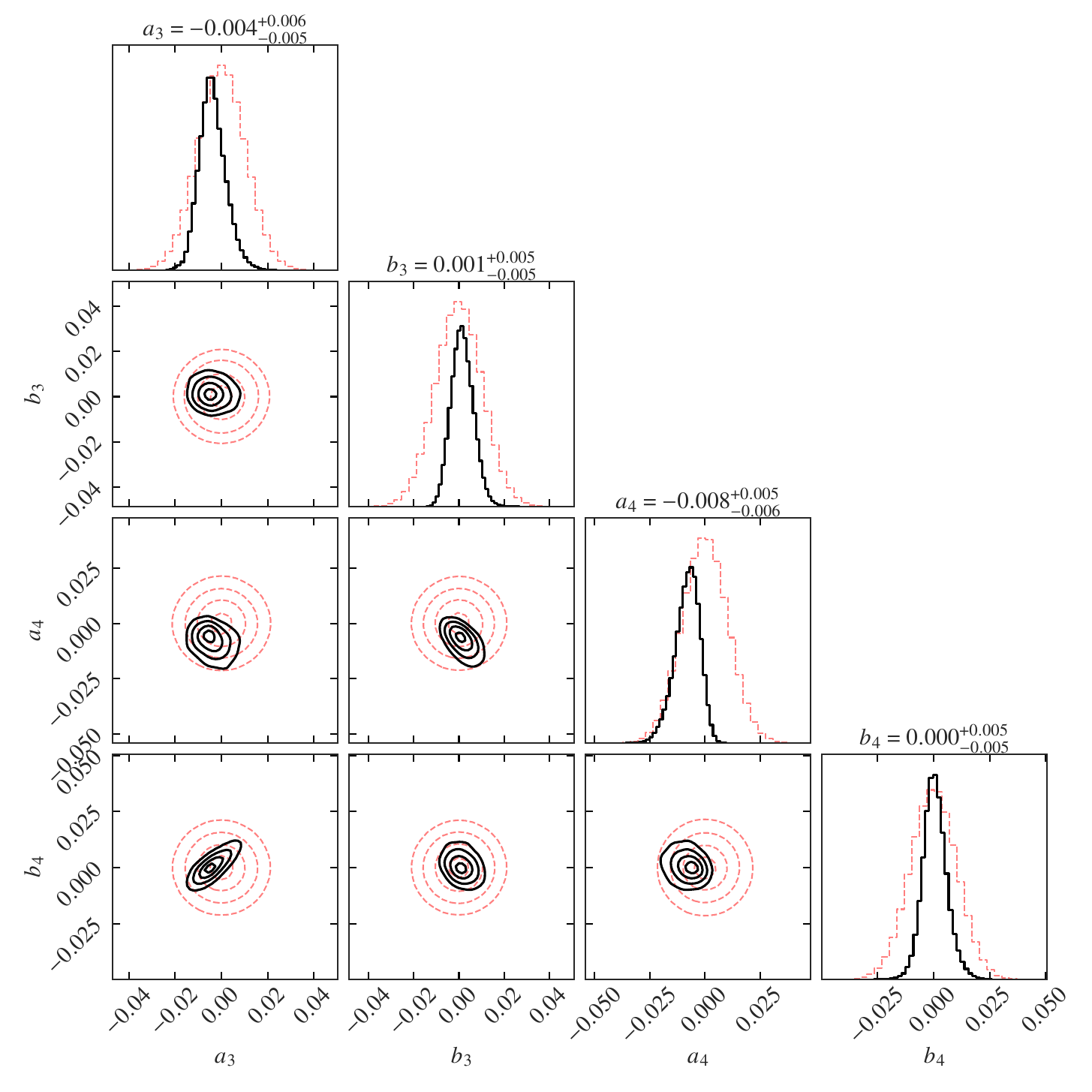}
    \label{fig:M1_eta_phi}
\end{subfigure}
\hfill
\begin{subfigure}{\columnwidth}
    \includegraphics[width=\columnwidth]{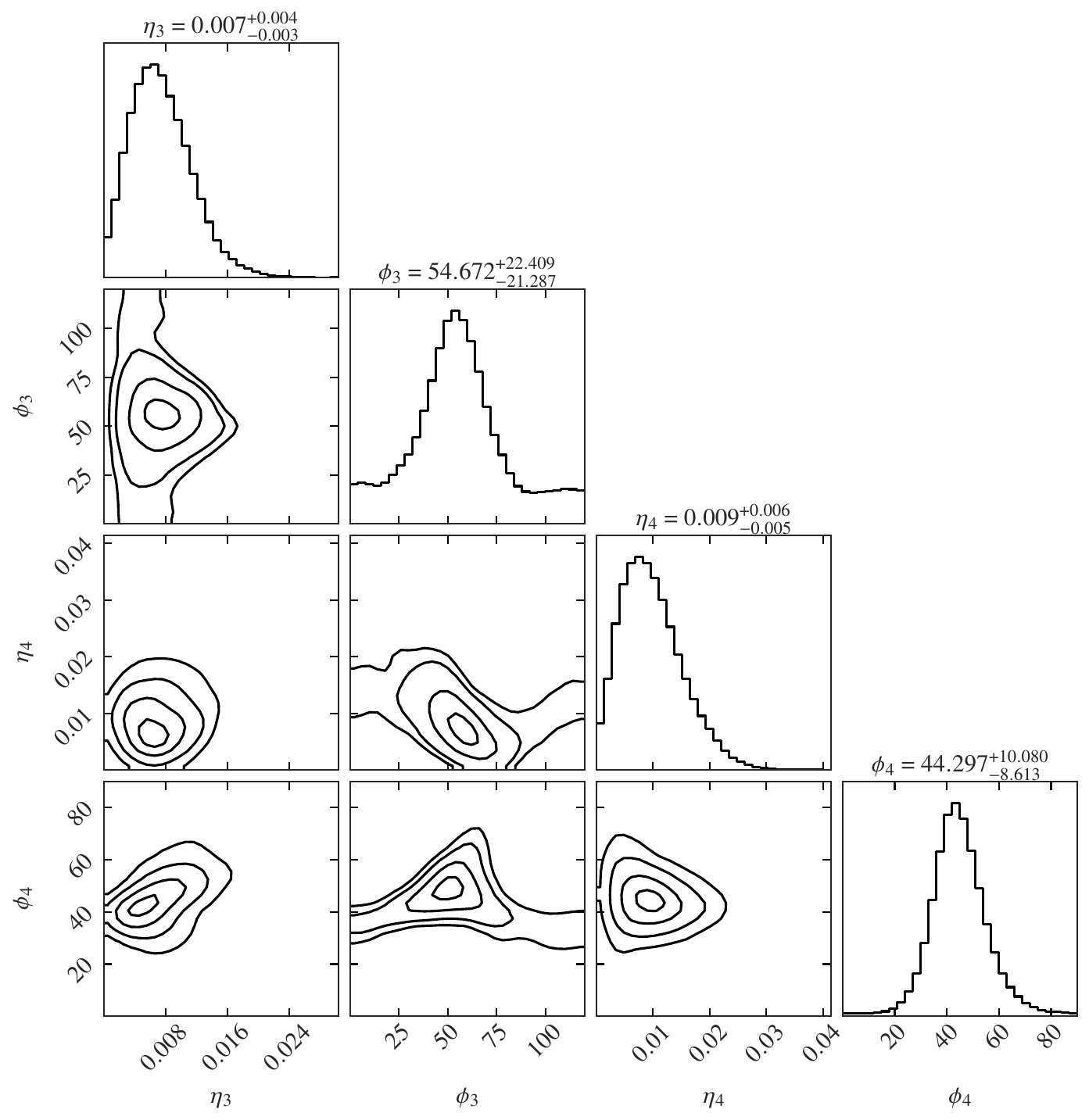}
    \label{fig:M1_amps}
\end{subfigure}
\caption{Same description as Figure \ref{fig:fig4}, except for the fold system.}
\label{fig:fig5}
\end{figure*}

\begin{figure*}
\centering
\begin{subfigure}{\columnwidth}
    \includegraphics[width=\columnwidth]{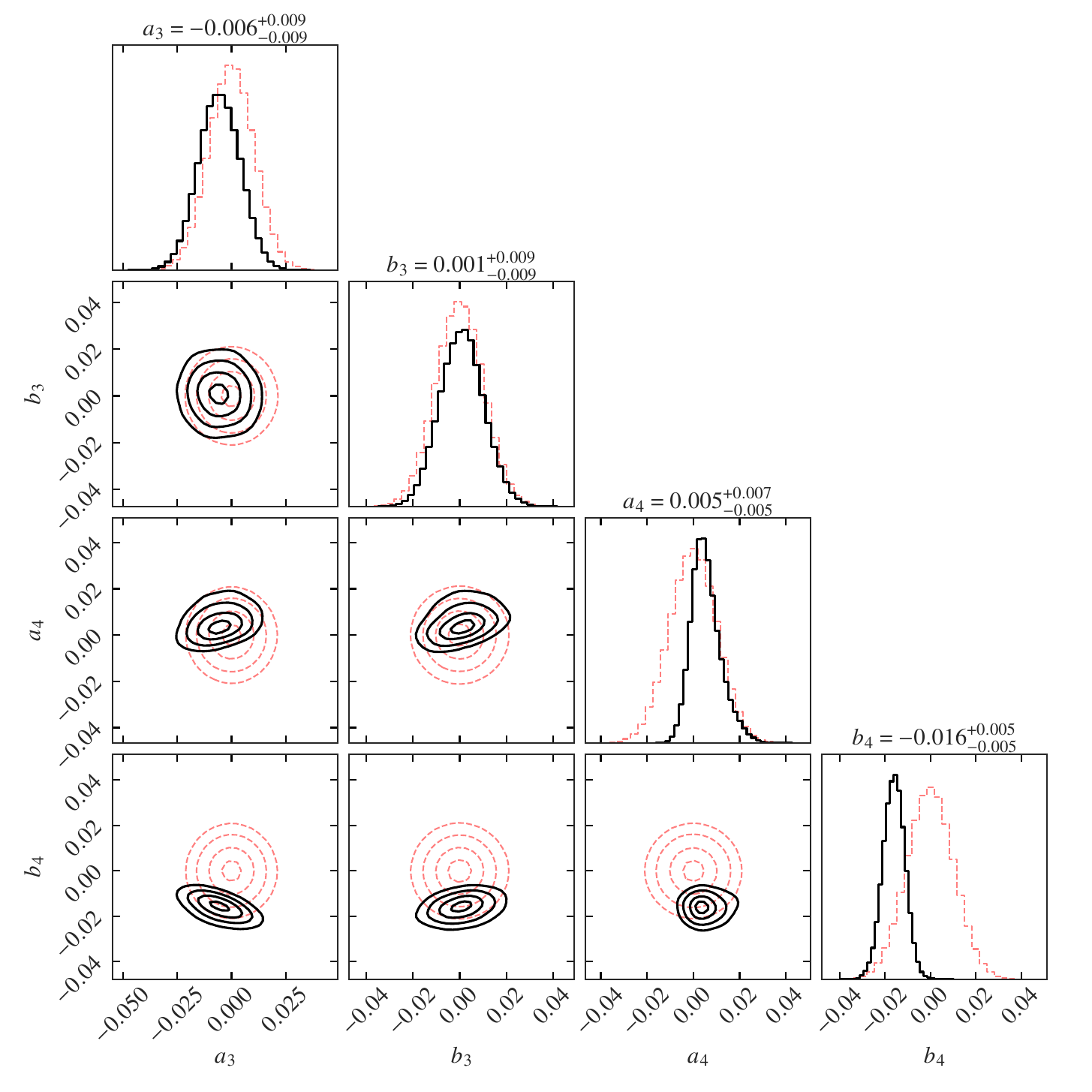}
    \label{fig:B1422_N_eta_phi}
\end{subfigure}
\hfill
\begin{subfigure}{\columnwidth}
    \includegraphics[width=\columnwidth]{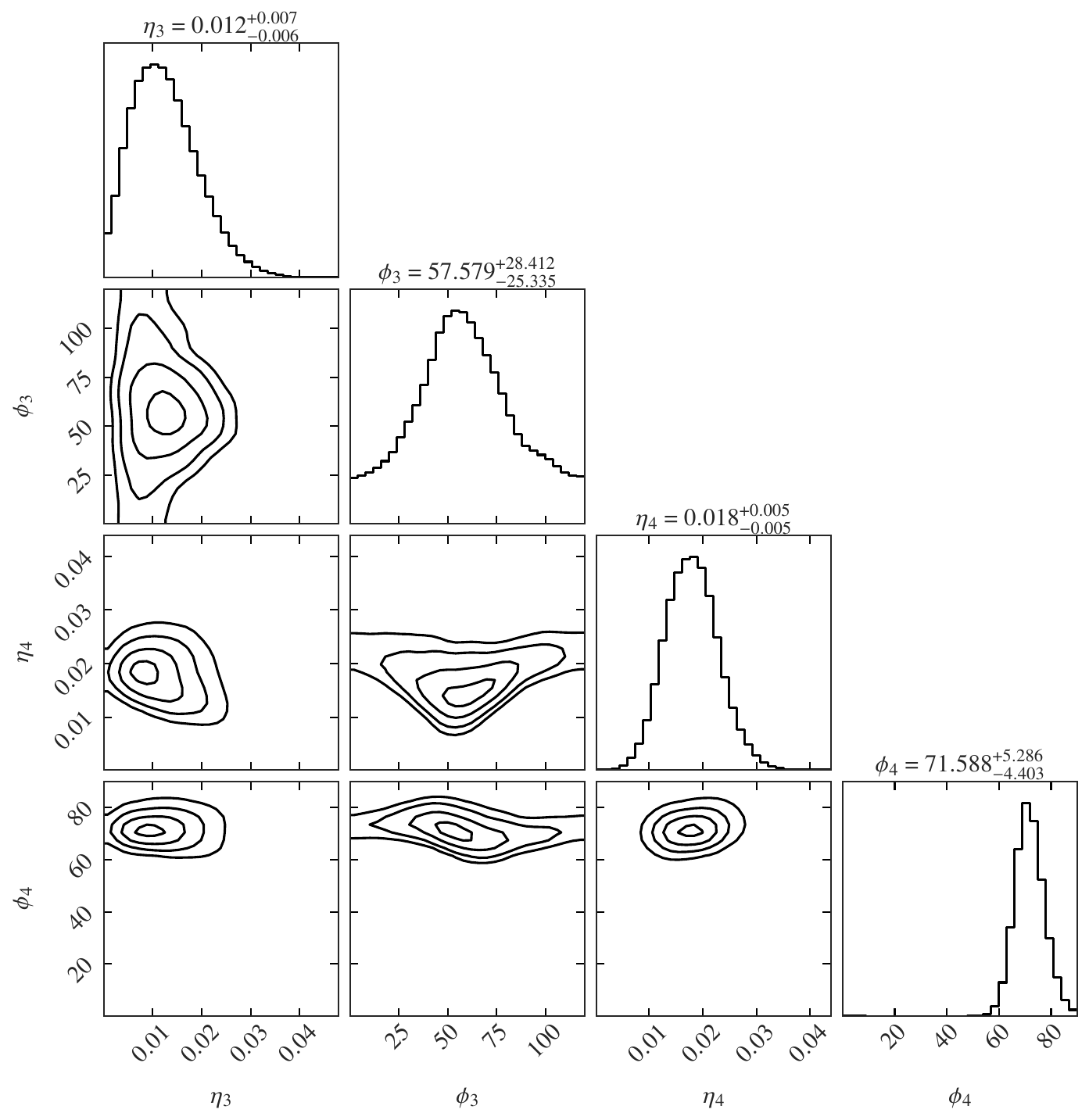}
    \label{fig:B1422_N_amps}
\end{subfigure}
\caption{Same description as Figure \ref{fig:fig4}, except for the cusp system.}
\label{fig:fig6}
\end{figure*}

\begin{figure*}
\centering
\begin{subfigure}{\columnwidth}
    \includegraphics[width=\columnwidth]{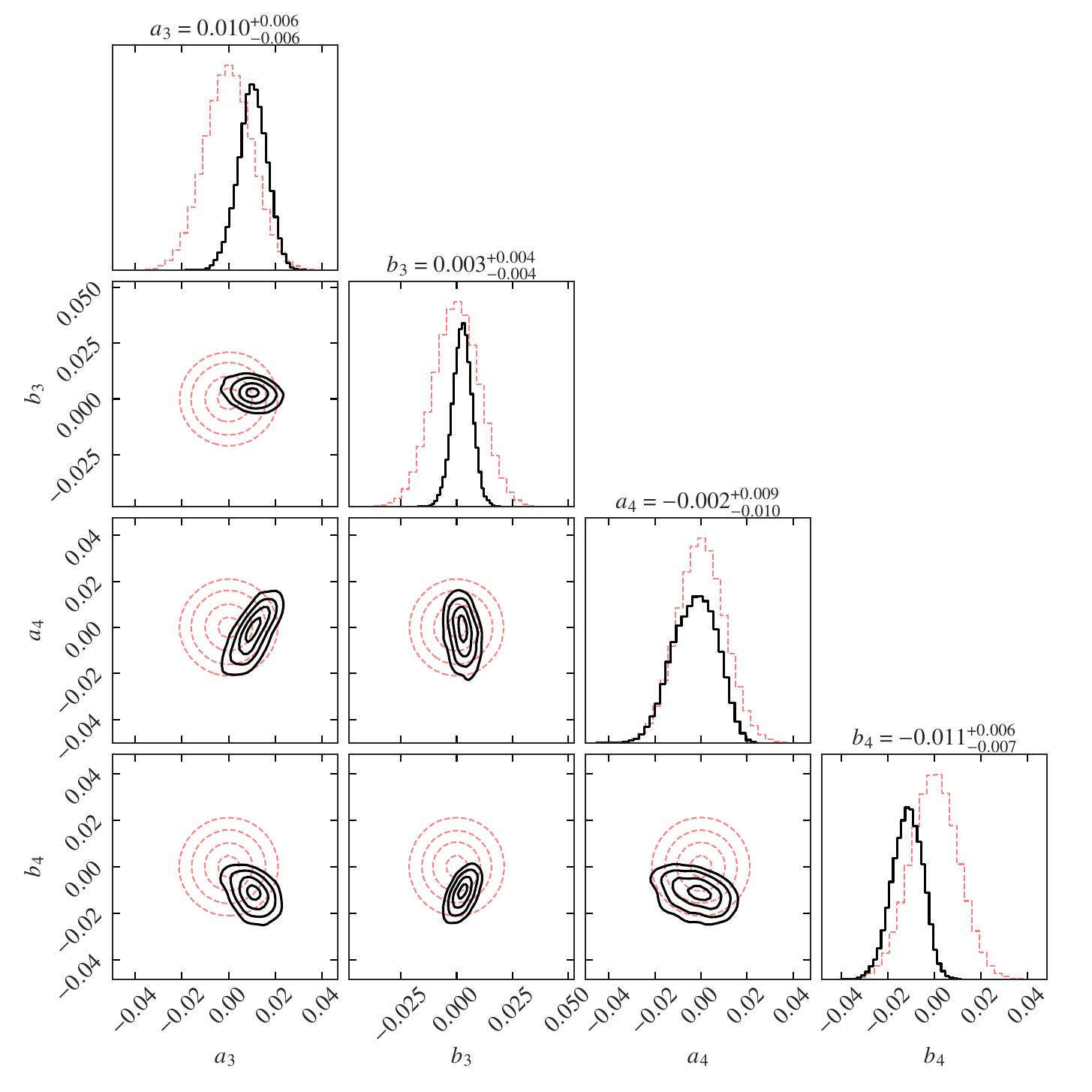}
    \label{fig:S1_eta_phi}
\end{subfigure}
\hfill
\begin{subfigure}{\columnwidth}
    \includegraphics[width=\columnwidth]{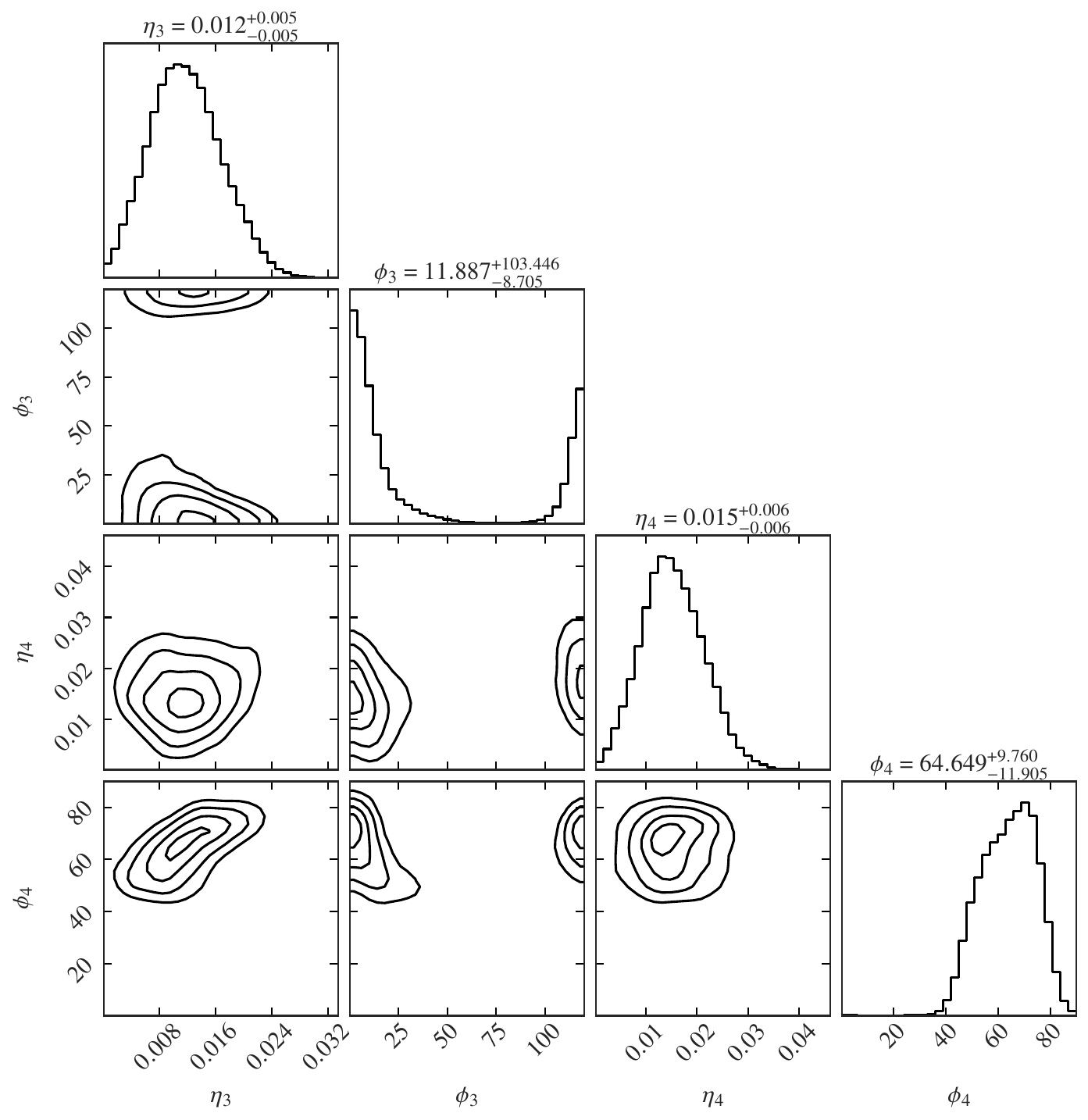}
    \label{fig:S1_amps}
\end{subfigure}
\caption{Same description as Figure \ref{fig:fig4}, except for the satellite system.}
\label{fig:fig7}
\end{figure*}

\begin{figure*}
\centering
\includegraphics[width=\textwidth]{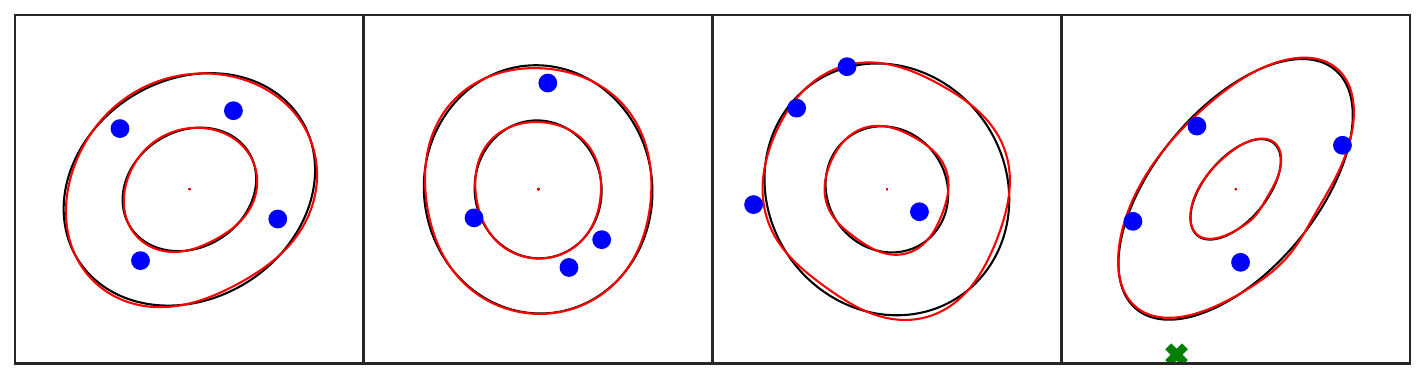}
\caption{Isodensity contours of the best-fitting models from the MCMC analysis of the cross, fold, cusp and satellite systems (left to right). The red curves show contours for the total convergence, including the EPL, shear and third- and fourth-order multipoles. The underlying black curves show contours of the EPL and shear only. Multipole parameter values: \textbf{cross}: $\eta_3=0.0109$, $\phi_3=93.91$, $\eta_4=0.0023$, $\phi_4=20.73$; \textbf{fold}: $\eta_3=0.0068$, $\phi_3=62.21$, $\eta_4=0.0052$, $\phi_4=44.86$; \textbf{cusp}: $\eta_3=0.0073$, $\phi_3=51.36$, $\eta_4=0.0151$, $\phi_4=73.19$; \textbf{satellite}: $\eta_3=0.0088$, $\phi_3=32.15$, $\eta_4=0.0037$, $\phi_4=54.78$.}
\label{fig:fig8}
\end{figure*}

With both third- and fourth-order multipole parameters free, all flux ratios in the simulated lens systems can be fit within 1$\sigma$. Figure \ref{fig:fig3} displays flux ratio histograms from the joint MCMC sampling of both multipole orders and macro-model parameters (\eplg$+$MP). For reference, we also include the flux ratio histograms that result from the MCMC sampling with only an \eplg. These \eplg-only models are unable to reproduce the simulated flux ratios within observational uncertainty in every case. Generally, the addition of third- and fourth-order multipoles to the model makes the model flux ratio distributions narrower and roughly centered around the simulated values. In other words, when multipoles with amplitudes that are consistent with the amplitudes of observed isophotes are included in the model, the simulated flux ratios do not appear anomalous despite the fact that they were perturbed by a population of CDM low-mass haloes, and, in one case, also a satellite galaxy. 

Figures \ref{fig:fig4}, \ref{fig:fig5}, \ref{fig:fig6} and \ref{fig:fig7} show the resulting posterior distributions for the third- and fourth-order multipole parameters for all four lens systems. Non-zero third- and fourth-order multipole amplitudes are preferred in every case, and most orientation angles are constrained within 20 degrees. 

Isodensity contours for the best-fitting models for each system and their underlying macro-models are shown in Figure \ref{fig:fig8}. The shapes of the contours for the cross and fold systems generally become 'boxier'. However, offsets from perfect alignment between the fourth-order multipole and the EPL major axes along with the inclusion of third-order multipoles, makes this effect more irregular than pure 'diskiness' or 'boxiness'. Deviation between the perturbed and unperturbed contours for the satellite system occurs roughly in alignment with the location of the missing satellite galaxy. It is worth noting again that the simulated data were generated without any multipoles.

\section{Discussion and conclusions}
\label{sec:conc}

We find that there is a substantial degeneracy between the perturbative lensing effects of low-mass dark matter haloes and multipoles when modelling quadruply-imaged quasars. Individually, third- and fourth-order multipoles with amplitudes around or below 0.01 can produce flux-ratio perturbations over 40 per cent. This is sufficient to reproduce perturbations induced not only by a population of low-mass CDM haloes, but also a satellite galaxy. Our results agree well with early investigations of this topic by \cite{evans2003} and \cite{cogndon2005}. From the analysis of real data, they find that all but the most extreme flux-ratio anomalies, such as those in the cusp systems B2045+265 and B1422+231, which differ from \eplg\ predictions in excess of $\sim$50 per cent, can be fit using multipoles with amplitudes smaller than 0.01. 

\subsection{A more general treatment of multipoles}

In our analysis of the ability of multipole components to reproduce flux ratio anomalies, we utilized two important general treatments of the multipoles: we included third-order multipoles in addition to the more standard fourth-order one, and we did not fix their position angle to the major axis of the underlying EPL model.

The freedom allowed to the position angle plays a key role in the degeneracy between multipoles and low-mass haloes. As we have shown, the orientations of multipoles that most closely reproduce the perturbed flux ratios in our simulated data do not always align with the EPL. \citet{evans2003}, \citet{kochanek2004} and \citet{cogndon2005} also found that a larger portion of observed flux-ratio anomalies can be reduced or eliminated when multipole angles are allowed to vary freely as opposed to being fixed.

Recent flux-ratio analyses, where fourth-order multipoles have been included \citep[e.g.][]{gilman2021, gilman2022, gilman2023, gilman2024}, have fixed the orientation angles to align with the EPL major axis. However, neither the total mass density nor the light distribution of real galaxies seem to suggest this assumption to be generally valid. Using high-angular resolution observations of real gravitational lens systems with an extended source, \cite{powell2022} and \cite{stacey2024} have found that the multipoles can be misaligned with respect to the underlying EPL. Additionally, tilted dark matter haloes are commonly observed in numerical simulations \citep[e.g][]{Han_2023}. At the same time, the observed isophotes of elliptical galaxies only have order $m=4$ perturbations aligned with the EPL for larger values of $a_4$, in excess of one per cent, i.e. for truly boxy isophotes \citep[][their fig. 4]{hao2006}. The fourth-order multipole in the best-fitting model for our cusp system has an amplitude of 1.5 percent, and it does not align with the EPL. The best-fitting models for our other systems all have fourth-order multipoles with amplitudes $\lesssim 0.5$ per cent. 

\subsection{Isophotes and iso-density contours}

An important comparison to our work is the study of isophotes in elliptical galaxies, since multipole patterns in the former are often used as motivation for including multipole components in the mass distribution of lens galaxies.

The amplitudes of the multipoles  that we find to be highly degenerate with low-mass haloes are consistent with amplitudes of the isophotes observed in elliptical galaxies. Indeed, our best-fitting values fall within the central 1-2 $\sigma$ of the distributions found by \cite{hao2006}, even though we use Gaussian priors that allow for more extreme values. Interestingly, despite isophote observations indicating that third-order multipoles may on average be weaker than fourth-order \citep{hao2006}, our findings indicate that they can still have significant impacts on flux ratios. We repeated our analysis with much tighter priors on the multipole amplitudes, Gaussians with widths 0.35 per cent instead of 1 per cent, and found that the cross and fold simulated lens systems can still be fit with $\chi^2<2$. The best-fitting realizations for the cusp and satellite systems both had $\chi^2<4$, much better than the best-fitting \eplg\ models. Even with priors on the multipole amplitudes that are comparable to, and tighter in the case of $a_4$, those suggested from isophotes, multipoles can have a significant impact on the flux ratios.

Still, it is currently unclear how closely galaxy mass distributions resemble their observed isophotes. Recently, \citet{stacey2024} analysed a sample of three strong gravitational lens systems and inferred multipole amplitudes in the total mass density distribution of up to $\sim$0.01 for both third and fourth orders. They found that in some cases the multipole amplitudes in the isophotes of the lens galaxies differ from those inferred in the total mass density distribution by 0.01 or more. Modelling very long baseline interferometric (VLBI) observations of a lensed radio jet, \cite{powell2022} found the third-order multipole to have an amplitude that is roughly twice as large as that of the fourth order, in disagreement with expectations from observations of isophotes in elliptical galaxies. Whether this is related to a genuine discrepancy between the light and mass distributions or to the fact that third-order multipoles in the mass density are more degenerate with low-mass haloes, at least for resolved lensing observations \citep{oriordan2023}, remains to be determined. As already pointed out by \citet{stacey2024}, the environment within which the lens galaxies reside may also play a role in observed differences between the isophotes and the iso-density contours. In general, our lack of knowledge about how well the projected mass density profile at the Einstein radius is traced by light limits our capability to use isophote distributions as a strong prior for the properties of multipoles in the mass distribution of gravitational lens galaxies.

\begin{figure}
\centering
\includegraphics[width=\columnwidth]{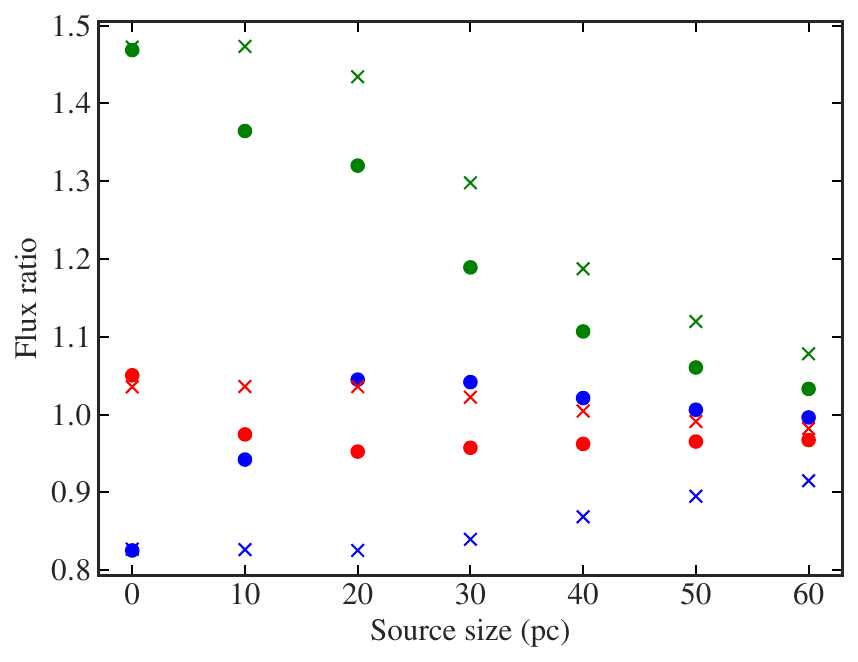}
\caption{Flux ratios as a function of source size for fixed \eplg+CDM (circles) and \eplg+M34 (crosses) models that predict the same image positions and point-source flux ratios. Source size denotes the standard deviation of a Gaussian brightness distribution. Note that the physical extents of the radio-emitting, mid-infrared-emitting and warm dust regions of quasars are typically expected to be $\lesssim 10$pc.}
\label{fig:fig9}
\end{figure}

\subsection{Considerations on the source size}

To understand the role played by the source on the relative effect of low-mass haloes and multipoles, we quantify the level of flux-ratio anomaly induced by the two contributions, i.e., low-mass haloes and multipoles, independently as a function of the source size (see Figure \ref{fig:fig9}). We select two models, one perturbed by low-mass haloes and the other by multipoles, which result in the same image positions and flux ratios from a lensed point source. As we increase the source size from a point to 40~pc, we find that the flux ratios perturbed by low-mass haloes approach their unperturbed values (i.e. the values expected from an \eplg-only mass distribution). Even when the source is only 10~pc in size, there is a $\sim$10 percent change of two of the flux ratios from their point-source values. On the other hand, the flux ratios perturbed by multipoles are hardly affected by increasing the source size up to 20 pc (i.e. 0-3 per cent). While in both cases larger source sizes produce smaller flux ratio anomalies, the effect is much smaller for the multipoles than for low-mass haloes. As the multipoles are less sensitive to changes in the source size, the more extended the source, the lower the multipole amplitudes that are needed to reproduce the effect of a given halo population. We can conclude, therefore, that our assumption of a point source gives a conservative quantification of the degeneracy between low-mass haloes and multipoles. 

\subsection{Implication for dark matter inferences using flux ratios}

Traditionally, flux-ratio constraints have been based on the concept that any anomaly that is seen with respect to the flux ratios predicted by the smooth \eplg\ model \citep[perhaps with the addition of a stellar disc;][]{hsueh2016, hsueh2017, hsueh2018}, is due to perturbations by clumpy dark matter. Thus, the underlying assumption is that the prevalence of observed flux-ratio anomalies can be used to infer the mass-function of low-mass dark matter haloes and, in turn, the properties of dark matter. If, instead, some unknown fraction of those anomalies are caused by multipole structure in the primary lensing galaxy, the constraining power of this method is diminished.

In this paper we have shown that, when presented with a simple set of observables, namely four image positions and four fluxes, it is generally possible to reproduce those observations with a model consisting of an \eplg\ base model plus either (1) low-mass CDM haloes but no multipoles or (2) multipoles but no low-mass haloes.  
The fluxes in real lens systems are likely to be affected by a combination of these two effects.
Because we cannot identify from just the lensed image positions and fluxes whether observed anomalies are due to low-mass haloes or multipoles, any dark matter inference from real flux-ratio data will be affected by what weight is given to each of these two components, i.e., by the priors on the relative contributions of the perturbing haloes and multipole components. For example, neglecting the contribution of the multipole component or reducing its complexity, e.g., by not including third-order multipoles or by fixing the multipole position angle to match the major axis of the EPL, will lead to a bias in favour of colder dark matter models. Most extremely, an inference could be made under the assumption that all perturbations were due to low-mass haloes, as was done in \cite{hsueh2020}, in which case a CDM-like model, which predicts large numbers of low-mass haloes, is likely to be preferred.  Similarly, an inference could be made assuming the maximum multipole contribution, in which case models that predict fewer low-mass haloes, such as WDM, are likely to be preferred.

All but the most extreme flux-ratio anomalies seen in real lensed quasar systems are comparable to or smaller than the anomalies considered in this paper, indicating that our ability to fit the anomalies with multipole components is not due to selectively choosing easy systems to fit. Thus, one conclusion to be drawn from our results is that, in the absence of other data that can inform our priors, lensed quasar flux ratios alone are ineffective at placing constraints on dark matter models.
 
In a follow-up paper, we will quantify how the inclusion of multipole components in the lens mass distribution affect existing constraints on dark matter from the two recent investigations by \cite{hsueh2020} and \cite{gilman2020}. Despite their different approach to multipoles, both works found nearly identical constraints on the WDM particle mass. However, neither work allowed for a general treatment of multipoles as presented in this paper. We expect that allowing more freedom in the angular structure of the lens galaxies will reduce the contribution from low-mass haloes to the observed flux ratios, and potentially weaken the constraints on dark matter.

\subsection{Future prospects}

As the prospects to derive meaningful constraints on the properties of dark matter with flux ratio only observations strongly depends on independent knowledge about the general mass density distribution of lens galaxies, it is fair to wonder where the required information could be obtained.

One option is deep observations that could allow one to identify and quantify the prevalence of multipoles in the light distribution of lens galaxies.  However, it is currently unknown how well the total mass distribution follows that of the light. Though the best-fit multipole amplitudes that we have found in this paper are not in tension with those observed in galaxy isophotes by \cite{hao2006}, the current number of quadruply-imaged quasar observations is not sufficient to precisely quantify the distributions of inferred multipole parameters across all elliptical galaxies from lens modelling.

Another option is modelling high-resolution images of strongly lensed extended emission, which is expected to allow for the constraint of more complex macro model features including multipoles \citep[e.g.][]{powell2022, stacey2024}. In principle, this information could be applied to the analysis of flux ratios when available \citep{gilman2024}. However, high-resolution imaging of lensed extended emission is at the moment not available for the majority of the $\sim 15$ lens systems in the current flux ratio sample. In addition, \cite{oriordan2023} have shown that the degeneracy between multipoles and low-mass haloes persists in the modelling of lensed extended emission, leaving as reliably detectable only low-mass haloes which are located in projection close to the lensed images. Furthermore, \cite{Herle2023} have shown that automatic lens finding techniques may lead to the selection of gravitational lens galaxies that have different mass distributions when the sources are extended instead of unresolved. If confirmed, their result may place some doubts on the possibility of transferring knowledge from one type of lens system to another.

\cite{galan2022} have recently shown that it may be possible to distinguish between the lensing effects of multipoles and low-mass haloes using extended sources and a non-analytical description of the lensing potential, but this method has yet to be demonstrated on real data. If proven successful, in combination with the large number of gravitational lens systems expected from the James Webb Space Telescope, Euclid and the SKA \citep{nierenberg2023, collett2015, mckean2015}, it may represent a path forward to obtain the necessary information on the mass density distribution of lens galaxies. Similarly, numerical simulations could provide useful insights on the mass and light distribution of (lens) galaxies 
\citep[e.g.][]{gao2006,lokas2022,despali2022b,Han_2023}.

To conclude, our results as well as previous similar works demonstrate that the analysis of flux-ratio anomalies, in its current form, cannot directly constrain the properties of the low-mass dark matter halo population. This follows from the main issue identified in this work, that the priors imposed on multipole parameters play a dominant role in the inference process. Though our focus is on CDM, we expect our results to be applicable to other dark matter models, but potentially with some differences. We expect, for example, the degeneracy here identified to be stronger for warm (WDM) and weaker for self-interacting (SIDM) dark matter. In the former case, the amplitude of the multipoles should not be significantly different than in CDM, but as low-mass haloes are less numerous and concentrated they result in weaker flux ratio anomalies. In SIDM models, galaxies are predicted to be rounder (\citealp[e.g.][]{brinckmann2018}; however, see \citealp{despali2022b}), while low-mass haloes, which experience core-collapse, have a stronger lensing effect \citep{gilman2023}. We will quantify the degeneracy between multipoles and low-mass haloes in different dark-matter models in a follow-up publication.

\section*{Acknowledgements}

J.~S.~C.\ and C.~D.~F.\ acknowledge support for this work from the National Science Foundation under Grant No. AST-1715611. S.~V. thanks the Max Planck Society for support through a Max Planck Lise Meitner Group. The authors thanks Hannah Stacey and John McKean for useful feedback and discussions. 

\section*{Data Availability}

The data used in this paper are available from the corresponding author on request.

\bibliographystyle{mnras}
\bibliography{multipole} 

\bsp	
\label{lastpage}
\end{document}